\documentclass[aps,pra,reprint,showkeys,longbibliography]{revtex4-2} 
\usepackage{amsmath}
\usepackage{amssymb}
\usepackage{graphicx} 
\usepackage{bm} 
\usepackage{xcolor}
\usepackage[toc,page]{appendix}
\usepackage{braket}
\usepackage{mathrsfs}

\makeatletter

\@ifclasswith{revtex4-2}{draft}{%
    \usepackage[notref,notcite]{showkeys}
    
}{}

\makeatother

\newcommand{\nn}{\nonumber}

\begin{document}

\title{Quantum field theory for classical fields}

\author{Christof Wetterich}
\affiliation{Institut f\"ur Theoretische Physik\\
    Universit\"at Heidelberg\\
    Philosophenweg 16, D-69120 Heidelberg}

\begin{abstract}
For classical field theories with probabilistic initial conditions the classical field observables are an idealization.
Their arbitrarily precise values poorly reflect the characteristic uncertainty in the presence of substantial fluctuations.
We propose to employ observables based on fluctuating fields.
In terms of these "statistical observables" the probabilistic classical field theory becomes a quantum field theory.
Non-commuting operators are associated to observables.
The quantum rules follow from the laws for classical probabilities.
The ``quantum'' is part of the ``classical''.
A regularized functional integral guarantees the unitarity of the quantum field theory. 
We discuss in detail the classical relativistic Klein-Gordon equation with interactions.
\end{abstract}

\maketitle


\indent We consider probabilistic classical field theories. 
A deterministic field equation governs the time evolution of the classical fields, while the initial conditions are given by 
a probability distribution.
The time evolution of the probability distribution is described by the Liouville equation.
Many analogies to quantum systems have been found for this setting.
A powerful operator formalism based on the pioneering work of Koopman \cite{KOOP} and von Neumann \cite{VNEU} has provided many
insights into the dynamics of classical probabilistic systems \cite{KAN, MAMA, MAU, GOMA, NIC, NAKL, VOLO, BON, AME, KPM, ACKA, CHRU, KLE, MEZI, DW}.
The functional integral for classical statistics constructed in ref. \cite{GOZ1, GRT} shows many similarities to quantum field theories.
The evolution equation for the time-dependent effective action, which generates the equal time correlation functions, looks very
similar for classical and for quantum fields \cite{CWET1, CWET2}.
Modifications of the Liouville equation can describe quantum particles \cite{CWQP1, CWQP2} or quantum fields \cite{CWFT1, CWFT2}.

\indent In this note we propose a different look at the Liouville equation by the use of statistical observables which are adapted
to the uncertainty in a situation with substantial fluctuations \cite{CWQP1, CWPW}.
The fluctuating field reflects the fact that a sharp probability distribution for the classical field would broaden rapidly.
The intrinsic uncertainty in the fluctuating field reflects the roughness of the probability distribution for the time derivatives of the fields. 
More generally, statistical observables measure or incorporate properties of the probabilistic information.
Well known examples are temperature or pressure in statistical equilibrium.
Statistical observables do not take fixed values in the "microstates", which are in our case the classical field configurations.
Probabilities for a pair of statistical observables to have simultaneously fixed values do not exist.
No classical correlation function can be defined for such pairs.
In consequence, correlations for such pairs do not have to obey Bell's inequalities \cite{BEL1, CHSH}.
There is therefore no obstruction that a classical probabilistic theory describes quantum systems.
We establish here that probabilistic classical field theories are equivalent to a type of quantum field theories if appropriate fluctuating
field observables are used.

\indent Probabilistic classical field theories of the type considered here have been investigated analytically and numerically as ``classical approximations'' to some quantum field theory \cite{PSTA, DTS, YK, AS1, BUJA, AS2, ABW, CKH, FGL, GLV, BBG, EGW, BER}.
This is not the approach taken here.
In our setting the probabilistic classical field theory is the starting point, and we show that this \textit{defines} a type of quantum field theory.
The quantum emerges from the classical.
In the absence of interactions the emergent quantum field theory coincides with the quantum field theory for which the classical approximation is taken.
In the presence of interactions they are different.

\indent Our starting point is a second order classical field equation for a field $\sigma(t, \vec{x})$, written in the form
\begin{equation}
    \label{eq:A1}
    \partial_t \sigma (t, \vec{x}) = \pi (t,\vec{x})\,, \quad \partial_t \pi (t, \vec{x}) = F\left( \sigma (t, \vec{x})  \right)\,.
\end{equation}
Here $\vec{x}$ is a position in a space with arbitrary dimension $D$, and $\sigma$, $\pi$ can be multi-component.
We will focus later on a single real scalar field with quartic interaction,
\begin{equation}
    \label{eq:A2}
    F = \partial_x^2 \sigma - m^2\sigma - \frac{1}{2} \lambda \sigma^3\,,
\end{equation}
for which eq.~\eqref{eq:A1} is the relativistic Klein-Gordon equation for a scalar with mass $m$ and self-interaction $\lambda$.
For our purpose it is useful to cast the probabilistic information at a given time $t$ into the form of a classical wave
function $q(t, \sigma, \pi)$ \cite{CWQP1, CWIT, CWQFC}, which is the square root of the time-local probability distribution $w(t,\sigma, \pi)$,
\begin{equation}
    \label{eq:A3}
    w(t,\sigma, \pi) = q^2(t, \sigma, \pi)\,.
\end{equation}

\indent The time evolution of the wave function obeys the Liouville equation
\begin{align}
    \label{eq:A4}
    &\partial_t q  = -\hat{L} q\,, \nn \\
    & \hat{L} = \int_{\vec{x}} \left( \pi (\vec{x}) \frac{\partial}{\partial \sigma (\vec{x})} + F(\vec{x}) \frac{\partial}{\partial \pi(\vec{x})}  \right)\,,
\end{align} 
from which the Liouville equation for $w$ follows.
The wave function is at every time a real unit vector,
\begin{equation}
    \label{eq:A5}
    \int \!\tilde{\mathcal{D}}\sigma \int \!\tilde{\mathcal{D}}\pi  \,q^2(\sigma, \pi) = 1\,,
\end{equation}
guaranteeing the normalisation and positivity of the probability distribution.
Here $\int \!\tilde{\mathcal{D}}\sigma$ is a functional integral over $\sigma(\vec{x})$ at fixed time $t$, and $\int \!\tilde{\mathcal{D}}\pi$
incorporates suitable $2\pi$-factors.
The time evolution describes a rotation of this real vector.
The use of a real wave function constitutes an important difference to the approach by Koopman and von Neumann who employ a complex Hilbert space.
After the introduction of a complex structure the phases play an important role, in contrast to the Koopman-von Neumann wave function where they are redundant.

\indent An important advantage of the use of wave functions is the possibility to perform basis transformations.
We employ this for our definition of fluctuating fields.
A functional Fourier transform maps $q$ to a complex wave function $\psi$
\begin{equation}
    \label{eq:A6}
    \psi (\sigma, \zeta) = \int\! \tilde{\mathcal{D}}\pi \exp \left\{ i\int_{\vec{x}} \zeta(\vec{x}) \pi (\vec{x}) \right\} q(\sigma, \pi)\,,
\end{equation}
which obeys the important selection rule
\begin{equation}
    \label{eq:A7}
    \psi (\sigma, -\zeta) = \psi^* (\sigma, \zeta)\,.
\end{equation}
The fluctuating field $\varphi$ and its mirror field $\chi$ are defined by
\begin{equation}
    \label{eq:A8}
    \varphi = \sigma + \frac{\zeta}{2}\,, \quad \chi = \sigma - \frac{\zeta}{2}\,.
\end{equation}
The mirror field $\chi$ is related to $\varphi$ by time reversal, $\pi \rightarrow -\pi$, $\zeta \rightarrow - \zeta$, $\varphi \leftrightarrow \chi$, implying complex conjugation of $\psi$,
\begin{equation}
    \label{eq:A9}
    \psi^* (\varphi, \chi) = \psi (\chi, \varphi)\,.
\end{equation}

\indent For the computation of expectation values of an observable $A$ one associates to it an operator $\hat A$ and uses the quantum rule,
\begin{equation}
    \label{eq:A10}
    \langle A \rangle = \psi^\dagger \hat{A} \psi = \int \! \tilde{\mathcal{D}}\varphi \tilde{\mathcal{D}}\chi \, \psi^*(\varphi, \chi) \hat{A} \psi(\varphi, \chi)\,.
\end{equation}
The operators for the classical fields are
\begin{align}
    \label{eq:A11}
    &\hat{\sigma} (\vec{x}) = \frac{1}{2} \left( \varphi (\vec{x}) + \chi (\vec{x}) \right)\,, \nn \\
    &\hat{\pi}(\vec{x}) = -i \frac{\partial}{\partial \zeta (\vec{x})} = -\frac{i}{2} \left( \frac{\partial}{\partial \varphi (\vec{x})} - \frac{\partial}{\partial \chi (\vec{x})} \right)\,.
\end{align}
For the classical fields the quantum rule \eqref{eq:A10} holds in an arbitrary basis.
It follows from the definition of expectation 
values in classical statistics.
Inverting the Fourier transform one may verify that the expectation values of functions $f(\sigma, \pi)$, with operators 
$f(\hat\sigma, \hat\pi)$, are given by the classical statistical rule
\begin{align}
    \label{eq:A12}
    \langle f(\sigma, \pi) \rangle &= q^T(\sigma, \pi) f(\hat\sigma, \hat\pi) q(\sigma, \pi) \nn \\
    &= \int\! \tilde{\mathcal{D}}\sigma \tilde{\mathcal{D}}\pi\, f(\sigma, \pi) w(\sigma, \pi)\,.
\end{align}
The operators $\hat\sigma$ and $\hat\pi$ commute, $\left[ \hat\sigma (\vec{x}), \hat\pi (\vec{y})\right] =0$, as appropriate for 
classical observables.

\indent In the basis \eqref{eq:A6} one has $\hat \zeta = \zeta$ and finds
\begin{align}
    \label{eq:A13}
    &\langle \zeta (\vec{x}) \rangle =0\,, \nn \\
    &\langle \zeta^2 (\vec{x}) \rangle = -q^T \frac{\partial^2}{\partial \pi (\vec{x})^2} q = \int\! \tilde{\mathcal{D}}\sigma \tilde{\mathcal{D}}\pi \left( \frac{\partial q}{\partial \pi (\vec{x})} \right)^2\,.
\end{align}
Thus, $\langle \zeta^2 (\vec{x}) \rangle$ is a measure for the roughness of the probability distribution in $\pi$ at a given 
position $\vec{x}$.
This makes it manifest that $\zeta$, and therefore also $\varphi$ and $\chi$, are statistical observables.
In the basis \eqref{eq:A6} one has $\hat \varphi = \varphi$, $\hat \chi = \chi$.
The expectation values of the fluctuating field and the mirror field coincide with the classical field
\begin{equation}
    \label{eq:A13A}
    \langle \varphi (\vec{x}) \rangle = \langle \chi (\vec{x}) \rangle = \langle \sigma (\vec{x}) \rangle \,.
\end{equation}
The dispersion obtains, however, an additional contribution from $\langle \zeta^2 \rangle$,
\begin{equation}
    \label{eq:A14}
    \langle \varphi^2 \rangle - \langle \varphi \rangle^2 = \langle \sigma^2 \rangle - \langle \sigma \rangle^2 + \frac{1}{4} \langle \zeta^2 \rangle \,,
\end{equation}
with all expectation values taken at $\vec{x}$.
The fluctuating field incorporates the observation that for fluctuations in $\pi$ a sharp value of the field is an unrealistic idealisation.
The fluctuating field operator $\hat\varphi$ does not commute with the classical field operator $\hat\pi$
\begin{equation}
    \label{eq:A15}
    \left[ \hat\varphi (\vec{x}), \hat\sigma (\vec{y}) \right]=0\,, \quad \left[ \hat\varphi(\vec{x}), \hat\pi (\vec{y}) \right] = -\frac{i}{2} \delta(\vec{x}-\vec{y})\,.
\end{equation}
This reflects the fact that $\varphi$ is a statistical observable which does not take a fixed value for a given configuration of the
classical fields $\sigma$ and $\pi$.

\indent Eq.~\eqref{eq:A4} becomes the complex Schrödinger equation 
\begin{align}
    \label{eq:C1}
    i \partial_t \psi &= H_s \psi = -\int_{\vec{x}} \left( \frac{\partial}{\partial \sigma (\vec{x})} \frac{\partial}{\partial \zeta (\vec{x})} + \zeta(\vec{x})F(\sigma(\vec{x})) \right) \psi \nn \\
    &= - \int_{\vec{x}} \Bigg\{ \frac{1}{2} \left( \frac{\partial^2}{\partial \varphi(\vec{x})^2 }  - \frac{\partial^2}{\partial \chi (\vec{x})^2}\right)  \\
    &\hspace{5mm}+ \left( \varphi(\vec{x}) - \chi (\vec{x}) \right) F \left( \frac{\varphi(\vec{x}) + \chi(\vec{x})}{2} \right)  \Bigg\}\psi\,, \nn
\end{align}
with $H_s$ the Fourier transform of $-i \hat{L}$.
One infers the operators for time-derivatives of the fluctuating field,
\begin{equation}
    \label{eq:C2}
    \hat{\dot{\varphi}} (\vec{x}) = \hat{\eta} (\vec{x}) = i\left[H, \hat{\varphi}(\vec{x}) \right] = -i \frac{\partial}{\partial \varphi (\vec{x})}\,.
\end{equation}
In contrast to the operators for the classical field the ones for the fluctuating field obey the standard quantum commutation relation 
\begin{equation}
    \label{eq:C3}
    \left[ \hat{\varphi}(\vec{x}), \hat{\eta}(\vec{y}) \right] = i \delta \left( \vec{x}-\vec{y}\right)\,.
\end{equation}
The phases of $\psi$ are important for the evolution and for expectation values.
These central properties for quantum mechanics are not realized in the Koopman-von Neumann approach.
Since eq.~\eqref{eq:C1} is a linear equation the superposition principle for solutions holds.
With Probabilities quadratic in $\psi$ the interference effects of quantum mechanics emerge naturally.

\indent We will establish that the expectation values of all observables which are functions of the fluctuating field $\varphi(\vec{x})$, or 
the associated "momentum type" observable $\hat \eta (\vec{x}) = -i \partial/\partial \varphi (\vec{x})$, are precisely the ones of an associated quantum field theory.
For this purpose one constructs a functional integral such that the expectation values obey
\begin{align}
    \langle &A(\varphi) \rangle = Z^{-1} \int \! \mathcal{D}\varphi \mathcal{D}\chi A(\varphi) e^{iS_M (\varphi, \chi)}\,, \nn \\
    &Z = \int \! \mathcal{D}\varphi \mathcal{D}\chi e^{iS_M (\varphi, \chi)}\,.
    \label{eq:A16}
\end{align}
Here $\int \mathcal{D}\varphi$ integrates over fields $\varphi (t, \vec{x})$ at all $t$.
The Minkowski action $S_M$ is real and $\exp (iS_M)$ is complex, as characteristic for a quantum field theory,
\begin{equation}
    \label{eq:A17}
    S_M = \int_x \left\{ \partial_t \sigma \partial_t \zeta + \zeta F(\sigma) \right\}\,.
\end{equation}
For the functional integral the field variables $\sigma$, $\zeta$ are functions of time and position.
We combine $t$ and $\vec{x}$ into a $D+1$-dimensional coordinate $x$, and $\int_x$ includes a time integration.
The functional integral \eqref{eq:A16} \eqref{eq:A17} is well known as the classical approximation of a quantum field theory.
We demonstrate here that despite its classical origin it defines itself a new type of quantum field theory.

\indent The action $S_M$ changes sign under the exchange $\varphi \leftrightarrow \chi$ and we write
\begin{equation}
    \label{eq:A18}
    S_M = \tilde S (\varphi) - \tilde{S} (\chi) -S_{int} (\varphi, \chi)\,.
\end{equation}
With $\partial^{\mu} \varphi \partial_{\mu} \varphi = -(\partial_t \varphi)^2 + (\partial_{\vec{x}}\varphi)^2$ one finds for the example \eqref{eq:A2}
\begin{align}
    \label{eq:A20}
    &\tilde S (\varphi) = -\int_x \left( \frac{1}{2} \partial^{\mu} \varphi \partial_{\mu} \varphi + \tilde{V}(\varphi)   \right)\,, \nn \\
    &\tilde{V} (\varphi) = \frac{m^2}{2} \varphi^2 + \frac{\lambda}{16}\varphi^4 \,, \nn \\
    &S_{int} = \frac{\lambda}{8} \int_x \left( \varphi^3 \chi -\varphi \chi^3 \right)\,.
\end{align}
This is the functional integral for a relativistic quantum field theory for the fluctuating field and the mirror field.
A field transformation $\varphi = \cosh \gamma\,\varphi^\prime + \sinh \gamma\,\chi^\prime$, $\chi = \cosh \gamma\,\chi^\prime + \sinh \gamma\,\varphi^\prime$ leaves the form of $S_M$ invariant and scales $\lambda \to e^{2\gamma} \lambda$.

\indent We regularize here the functional integral in a way which demonstrates directly the unitarity of the quantum field theory.
We take a lattice of space-time points $x$ with lattice distance $\varepsilon$.
The classical field equation can be formulated as a cellular automaton \cite{VNEUA, ULA, ZUS, WOL, THO}, with a sequence of two updating steps,
\begin{align}
    \label{eq:A21}
    &\pi (t+\varepsilon, \vec{x}) = \pi(t, \vec{x}) + \varepsilon F(t,\vec{x})\,, \nn \\
    & \sigma (t+\varepsilon, \vec{x}) = \sigma(t, \vec{x}) + \varepsilon \pi (t+\varepsilon, \vec{x})\,.
\end{align}
For the example \eqref{eq:A2} one takes
\begin{align}
    \label{eq:A22}
    F(t,\vec{x}) = &\frac{1}{2\varepsilon^2}\sum_k \left( \sigma (t, \vec{x}+\varepsilon_k) + \sigma (t, \vec{x}-\varepsilon_k) -2\sigma (t, \vec{x}) \right) \nn\\
    &-m^2\sigma (t, \vec{x}) - \frac{1}{2} \lambda \sigma^3 (t,\vec{x})\,,
\end{align}
with $\varepsilon_k$ the lattice vector in the $k$-direction, $|\varepsilon_k| = \varepsilon$.
The updating of the fields at $\vec{x}$ is influenced only by the field values at $\vec{x}$, $\vec{x}-\varepsilon_k$ and $\vec{x}+\varepsilon_k$.
This cellular property induces the characteristic causal structure of quantum field theories with light cones.
For the updating from $t+\varepsilon$ to $t+2\varepsilon$ we reverse the order of the two steps in eq.~\eqref{eq:A21} in order to implement
invariance under time reflection.

\indent For a probabilistic automaton the value of the wave function $q(t, \sigma^\prime, \pi^\prime)$ for a particular field configuration
$(\sigma^\prime, \pi^\prime)$ is transported at $t+2\varepsilon$ to the field configuration $(\sigma, \pi)$ which obtains from  $(\sigma^\prime, \pi^\prime)$
by the updating.
This can be described by a step evolution operator $\hat S (t)$,
\begin{align}
    \label{eq:A23}
    &q(t+2\varepsilon)  =\hat{S}(t) q(t)\,, \\
    &q(t+2\varepsilon; \sigma, \pi) = \int \! \tilde{\mathcal{D}}\sigma^\prime \tilde{\mathcal{D}}\pi^\prime \hat{S}(t; \sigma, \pi, \sigma^\prime, \pi^\prime) q(t; \sigma^\prime, \pi^\prime)\,, \nn
\end{align}
with
\begin{align}
    \label{eq:A24}
    \hat S = &\delta \left( \sigma-\sigma^\prime - \varepsilon(\pi+\pi^\prime) + \varepsilon^2 (F(\sigma)- F(\sigma^\prime) ) \right) \nn \\
    &\times \delta \left( \pi-\pi^\prime -\varepsilon (F(\sigma) + F(\sigma^\prime)  ) \right)\,.
\end{align}
Eq.~\eqref{eq:A23} encodes the updating \eqref{eq:A21} from $t$ to $t+\varepsilon$, and the subsequent one from $t+\varepsilon$ to $t+2\varepsilon$.
Taking the continuum limit $\varepsilon \rightarrow 0$ it reproduces the Liouville equation \eqref{eq:A4}.

\indent The overall probability distribution $\tilde{w}[\sigma, \pi]$ allows one to compute observables for all $t$ and $\vec{x}$.
The arguments are the fields $\sigma(x)$, $\pi(x)$ with $x=(t,\vec{x})$.
Expectation values for classical field observables read 
\begin{equation}
    \label{eq:A25}
    \langle A[\sigma, \pi] \rangle = \int\! \mathcal{D}\sigma \int\! \mathcal{D}\pi \,\tilde{w}[\sigma, \pi] A[\sigma, \pi]\,,
\end{equation}
with functional integrals over $\sigma(x)$, $\pi(x)$.
We construct $\tilde{w}$ by a sequence of step evolution operators in time steps $2\varepsilon$,
\begin{align}
    \label{eq:A26}
    &\tilde{w}[\sigma, \pi] = q(t_f) \prod_{t=t_{in}}^{t_f - 2\varepsilon} K(t)\, q(t_{in})\,, \nn \\
    &K(t) = \hat{S} \left( \sigma(t+2\varepsilon), \pi(t+2\varepsilon); \sigma(t), \pi(t) \right)\,.
\end{align}
The boundary terms are given by the wave functions at initial and final time $t_{in}$ and $t_f$.
Those are functions of $\sigma(t_{in})$, $\pi(t_{in})$ and $\sigma(t_f)$, $\pi(t_f)$, respectively.
By virtue of the $\delta$-functions in $\hat S$ the overall probability distribution is non-zero only for the trajectories in field
space which are allowed by the updating.
Each trajectory is specified by a field configuration at $t_{in}$, and its probability equals $q^2(t_{in})$ for this configuration.
We can write $\tilde w$ in terms of a real action $S$,
\begin{equation}
    \label{eq:A27}
    \exp (-S) = \prod_t K(t) \,\,\, \text{"="}\,\,\, \delta(\sigma - \sigma_{cl}) \delta(\pi-\pi_{cl})\,,
\end{equation}
where $(\sigma_{cl}, \pi_{cl})$ are the solutions of the classical field equation with initial field configurations specified at $t_{in}$.
Similar $\delta$-functions appear in other constructions of functional integrals for classical field equations \cite{GOZ1, MSR, JAN, DOM, GRT, BER}.

\indent Performing the Fourier transform \eqref{eq:A6} changes $\hat S$ to $\hat{S}_s$, where we omit irrelevant constant factors,
\begin{align}
    \label{eq:A28}
    &\hat{S}_s (\sigma, \zeta, \sigma^\prime, \zeta^\prime) = \exp \left( -2i\varepsilon H_s \right)\,, \\
     &H_s = -\int_{x} \left[  \frac{(\zeta-\zeta^\prime)(\sigma-\sigma^\prime)}{4\varepsilon^2}+\frac{1}{2}\left( \zeta F(\sigma) + \zeta^\prime F(\sigma^\prime) \right)\right]\,. \nn
\end{align}
Correspondingly, this changes $\exp (-S) \rightarrow \exp (iS_M)$, and one recognizes eq.~\eqref{eq:A17} in the continuum limit $\varepsilon \rightarrow 0$.
The (unnormalized) overall weight distribution is given in this basis by
\begin{equation}
    \label{eq:A29}
    \tilde{w}_s = \psi^* (t_f) \exp (iS_M) \psi (t_{in})\,.
\end{equation}
(We have omitted the boundary term eq.~\eqref{eq:A16}.)
For our regularisation one has to establish Lorenz symmetry in the continuum limit.
One could choose different regularisations, for example by transforming to momentum space and employing an invariant momentum cutoff.
Our regularisation based on a cellular automaton has the advantage of a straightforward implementation for numerical simulations.

\indent For a probabilistic cellular automaton the evolution is manifestly unitary.
This provides for a direct proof that the regularized functional integral describes a unitary quantum field theory.
All characteristic features of quantum mechanics are realized.
The evolution follows a complex Schrödinger equation with a hermitian Hamiltonian.
The phases matter for the interference from a superposition of two solutions.
Operators for relevant observables do not commute, entailing the uncertainty principle.
The unusual feature is the particular form of the Hamiltonian with positive and negative eigenvalues.

\indent An important simplification occurs in the absence of interactions, $\lambda=0$.
In this limit the mirror field decouples from the fluctuating field.
For observables depending only on the fluctuating field $\varphi$ we can integrate out $\chi$.
This only induces an irrelevant constant in $\tilde{w}$.
Once we focus on observables depending only on $\varphi$ the probabilistic classical field theory is directly equivalent to a standard quantum field theory for a free scalar field.
This extends to the presence of external fields which modify the field equations while keeping them linear in $\sigma$ and $\pi$.

\indent In the presence of interactions for $\lambda>0$ we can still integrate out $\chi$.
The result depends on $\varphi$, however, and the properties of the reduced functional integral are not well known.
The reduced weight factor,
\begin{align}
    \label{eq:10}
    \bar{w}[\varphi] = \int\! \mathcal{D}\chi \tilde{w} [\varphi, \chi] &= \int\! \mathcal{D}\chi \exp \left\{ iS_M(\varphi, \chi) \right\} \nn \\
    &=  \exp \left\{ iS(\varphi) \right\}\,,
\end{align}
is sufficient for computing the expectation values of all observables which are functionals of $\varphi$.
(In the most general setting $S(\varphi)$ may depend on boundary terms.
For the example of direct product initial wave functions $\psi (\varphi, \chi) = \tilde{\psi} (\varphi) \tilde{\psi}^* (\chi)$
the integrand in eq.~\eqref{eq:10} receives an additional boundary factor $\tilde{\psi} (t_f, \chi) \tilde{\psi}^*(t_{in}, \chi)$,
and $\bar{w}(\varphi)$ involves a factor $\tilde{\psi}^* (t_f, \varphi) \psi (t_{in},\varphi)$.)
All observables constructed from the fluctuating fields obey the standard relation of quantum field theory 
\begin{equation}
    \label{eq:10A}
    \langle A(\varphi) \rangle = \int \! \mathcal{D}\varphi \hat{A}(\varphi) \bar{w}(\varphi)\,.
\end{equation}

\indent The integration over $\chi$ induces a correction to the action for $\varphi$
\begin{equation}
    \label{eq:A30}
    S(\varphi) = \tilde{S} (\varphi) + \Delta S(\varphi)\,,
\end{equation}
with
\begin{align}
    \label{eq:9}
    &\Delta S (\varphi) = -i \ln \bar{Z}^* (\varphi)\,, \quad \bar{Z}(\varphi) = \int\! \mathcal{D}\chi \exp\left\{ i \bar{S}(\varphi,\chi) \right\}\,, \nn \\
    &\bar{S}(\chi, \varphi) = \tilde{S}(\chi) + S_{int} (\varphi, \chi)\,. 
\end{align}
We identify $\bar{Z}[\varphi]$ as the partition function of a quantum field theory for $\chi$ with a source term $\sim \varphi^3$ and a cubic term
$\sim \varphi$.

\indent One may evaluate eq.~\eqref{eq:9} in a saddle point approximation.
In lowest order (tree approximation) this yields
\begin{equation}
    \label{eq:11}
    \Delta S^{(0)}(\varphi) = - \bar{S}(\varphi, \bar{\chi}(\varphi))\,,
\end{equation}
with $\bar{\chi}(\varphi)$ a solution of the field equation $\partial \bar{S}/ \partial \chi =0$,
\begin{equation}
    \label{eq:12}
    (m^2-\partial^{\mu} \partial_{\mu}) \chi = \frac{\lambda}{8} \left( \varphi^3 - 3\chi^2\varphi - 2\chi^3 \right)\,,
\end{equation}
resulting in 
\begin{equation}
    \label{eq:13}
    \Delta S^{(0)}(\varphi) = -\frac{\lambda}{16} \left(  \bar{\chi}^4 + \bar{\chi}^3 \varphi + \bar{\chi}\varphi^3  \right)\,.
\end{equation}
For slowly varying $\varphi$ and small $\lambda \varphi^2 / m^2$ the solution of eq.~\eqref{eq:12} corrects the potential \eqref{eq:A20}
\begin{align}
    \label{eq:14}
    &\bar{\chi} = \frac{\lambda \varphi^3}{8m^2} - \frac{9\lambda^3 \varphi^7}{512m^6} + \cdots \,, \nn \\
    &\Delta \tilde{V} =  \frac{\lambda^2 \varphi^6}{128 m^2} +  \cdots \,.
\end{align}
Taking into account derivatives replaces in momentum space $m^2$ by $m^2 + q^{\mu}q_{\mu}$, with $q_{\mu}$ the momentum of the $\varphi^3$-operator.
This is the standard result for the exchange of a massive particle.

\indent The next order in the saddle point expansion yields the one loop correction.
For constant $\varphi$ the $\varphi$-dependent action for $\chi$ reads in quadratic order in $\chi^\prime = \chi - \bar{\chi}$
\begin{equation}
    \label{eq:B1}
    \hat{S}_2 = -\frac{1}{2} \int_x \delta\chi \left( m^2+\frac{3 \lambda^2 \varphi^4}{32 m^2} - \partial^{\mu} \partial_{\mu} \right)\delta\chi\,.
\end{equation}
The Gaussian integration leads to a one-loop momentum integral, $d=D+1$,
\begin{align}
    \label{eq:B2}
    \Delta S^{(1)} &= -i \ln \int \! \mathcal{D} \chi^\prime \exp \left( -i\bar{S}_2 (\chi^\prime, \varphi) \right) \nn \\
    &= \frac{1}{2} \int\!\frac{d^d q}{(2\pi)^d} \ln\left( q^2 + m^2 + \frac{3\lambda^2 \varphi^4}{32 m^2} \right)\,,
\end{align}
which depends on the regularisation for large $q^2$.
For our discrete formulation it yields a finite contribution to the potential $\sim \lambda^2 \varphi^4/(m^2 \varepsilon^2) + \cdots$, which dominates
the correction for small $\varphi$.

\indent In the presence of interactions it is not obvious if $\varphi$ is an optimal choice for a quantum field.
One would rather like to find a pair of modified fields $\varphi^\prime$ and $\chi^\prime$ related by time reversal such that the Hamiltonian becomes block diagonal with opposite eigenvalues in the blocks.
The relation of such fields to $\sigma$ and $\pi$ may be non-linear.
A selection criterion for the physical quantum field could be the independence of its correlation functions from the ultraviolet cutoff, as for the renormalized field in quantum field theory.

\indent Our approach can be extended to general first order classical transport equations with probabilistic conditions \cite{CWTE}.
This permits to make contact with the functional integral of ref.~\cite{MSR, JAN, DOM}.
In case of a stochastic force one deals with a Fokker-Planck equation \cite{KAM}.

\indent Our overall conclusion is rather simple: Using fluctuating field observables a probabilistic classical field theory can be described by
the functional integral of a unitary quantum field theory.
In the absence of interactions a reduced quantum field theory can be defined for a subsystem.
It is of the standard type for a free quantum field theory.
This includes the presence of external fields which render $F$ in eq.~\eqref{eq:A1} inhomogeneous in space with explicit dependence on $\vec x$.
For example, a constant $m$ in eq.~\eqref{eq:A2} can be replaced by $m + V(\vec x)$.
For a complex scalar field $\sigma$ a conserved charge is an important statistical observable.
It allows for the construction of a vacuum state and particle excitations.
The one-particle state describes in the non-relativistic limit a standard quantum particle in an arbitrary potential $V(x)$ \cite{CWDS}.
In this sense quantum mechanics emerges from classical statistics.

\indent \textit{Acknowledgement:} The author would like to thank J. Berges and J. Marijan for stimulating discussions.

\nocite{*}
\bibliography{refs}

\end{document}